\documentstyle[12pt]{article}

\newcommand {\nn}    {\nonumber}
\newcommand {\vs}[1]  { \vspace*{#1 cm} }

\newcounter{eq}
\newcounter{sc}

\newcommand {\PL}   {Phys.Lett.}
\newcommand {\PR}   {Phys.Rev.}
\newcommand {\PRL}   {Phys.Rev.Lett.}

\newcommand {\JHEP}  {JHEP}

\def\overleftrightarrow#1{\vbox{\ialign{##\crcr
 $\leftrightarrow$\crcr\noalign{\kern-1pt\nointerlineskip}
 $\hfil\displaystyle{#1}\hfil$\crcr}}}

\newlength{\minitwocolumn}
\begin{document}


\begin{flushright}
\end{flushright}
\vspace{30pt}

\pagestyle{empty}
\baselineskip15pt

\begin{center}
{\large\bf Localization of Matters on a String-like
Defect

 \vskip 1mm
}

\vspace{20mm}

Ichiro Oda
          \footnote{
          E-mail address:\ ioda@edogawa-u.ac.jp
                  }
\\
\vspace{10mm}
          Edogawa University,
          474 Komaki, Nagareyama City, Chiba 270-0198, JAPAN \\

\end{center}


\vspace{15mm}
\begin{abstract}
After presenting string-like solutions with a warp factor
to Einstein's equations,
we study localization of various spin fields on a string-like
defect in a general space-time dimension from the viewpoint of
field theory. 
It is shown that spin 0 and 2 fields are localized on a defect with 
the exponentially decreasing warp factor. 
Spin 1 field can be also localized on a defect 
with the exponentially decreasing warp factor.
On the other hand, spin one-half and three-half fields can be 
localized on a defect with the exponentially increasing warp 
factor, provided that additional interactions are not introduced. 
Thus, some mechanism of localization must be 
invoked for these fermionic fields.
These results are very similar to those of a domain wall in five 
space-time dimensions except the case of spin 1 field.

\vspace{15mm}

\end{abstract}

\newpage
\pagestyle{plain}
\pagenumbering{arabic}


\rm
\section{Introduction}

It is nowadays believed that the idea of extra dimensions 
would be one of the most intriguing ideas concerning unification
of gauge fields with general relativity.
For instance, in superstring theory, the traditional 
method of the compactification involves writing a ten dimensional
manifold as a direct product of a non-compact four dimensional
Minkowski space-time and a compact six dimensional Ricci flat manifold 
(or non-manifold like an orbifold) with the size of the compact space
being set by a string scale. Here it is usually assumed that the size
of extra spatial dimensions is so small that it is not observed as yet.
An important point in this method is that one needs
some unknown mechanism of stabilizing the size of all six extra
dimensions via non-perturbative string effects. However, it is
well known that stabilizing the moduli associated with the size
of extra spatial dimensions is a very difficult and unsolved problem
\cite{Green, Pol}, which would be a great shortcoming within
the present framework of superstring theory.
 
In recent years, an alternative scenario of the compactification 
has been put forward \cite{Randall2}. This new idea is based on  
the possibility that our world is a three brane embedded in
a higher dimensional space-time with non-factorizable warped
geometry. In this scenario, it is $\it{a \ priori}$ assumed that 
all the matter fields are constrained to live on the three brane, 
whereas gravity is free to propagate in the extra dimensions.
However, this assumption about trapping of Standard Model particles 
on the brane is not so obvious at first glance. 
In string theory, such particles
could be naturally localized on D3-brane \cite{Polchinski}, but
it is fair to say that string theory realization of the alternative 
compactification scenario
is not completely understood yet. Then it is quite of interest to
ask whether field theoretic localization mechanism works as
well or not. Indeed, it has been already known that not only scalars 
and fermions \cite{Rubakov} but also gauge fields \cite{Dvali} 
are localized on the brane in terms of field theoretic mechanism.

This localization mechanism has been recently investigated in $AdS_5$
space \cite{Rizzo, Pomarol, Grossman, Chang, Bajc}.  In particular, in
the Randall-Sundrum model in five dimensions \cite{Randall2}, 
the following facts are clarified: spin 0 field is localized
on a brane with positive tension which also localizes the graviton
\cite{Bajc}. Spin 1 field is not localized neither on a brane
with positive tension nor on a brane with negative tension
\cite{Rizzo, Pomarol}. Moreover, spin 1/2 and 3/2 fields are localized
not on a brane with positive tension but on a brane with negative
tension \cite{Grossman, Chang, Bajc}. Thus, in order to fulfill the
localization
of Standard Model particles on a brane with positive tension,
it seems that some additional interactions except gravity must 
be also introduced in the bulk. 

The aim of the present article is to extend these works to the case of 
a string-like defect with codimension 2 in a general space-time dimension.
In this respect, it is valuable to ask why we consider the object with 
codimension 2.  One reason is, of course, because it is 
in itself of interest to generalize many interesting properties obtained
in the Randall-Sundrum model in five dimensions to the string-like solutions
in general $D$ dimensions, with $D=6$ being of special importance to 
physics. But the most important reason comes from
the observation that logarithmic gauge coupling unification may be
achieved in theories with (sets of) two large spatial dimensions
\cite{Arkani1}. The logarithmic behavior of the Green's functions
in effectively two dimensions has a chance of giving rise to 
logarithmic variation of the parameters on our brane, thereby 
reproducing the logarithmic running of coupling constants.

The solutions to Einstein's equations in two extra dimensions
have been so thus studied by many groups \cite{Chodos, Cohen, Gregory, 
Cvetic, Csaki, Vilenkin, Gherghetta, Minic, Chaichian, Chodos2}.
In this article, we shall present slightly more general solutions with 
a warp factor in a general space-time dimension.

\section{Einstein's equations}

The action which we consider in this article is that of gravity
in general $D$ dimensions, with the conventional Einstein-Hilbert
action and some matter action:
\begin{eqnarray}
S = \frac{1}{2 \kappa_D^2} \int d^D x  
\sqrt{-g} \left(R - 2 \Lambda \right) 
+ \int d^D x  \sqrt{-g} L_m,
\label{1}
\end{eqnarray}
where $\kappa_D$ denotes the $D$-dimensional gravitational
constant with a relation $\kappa_D^2 = 8 \pi G_N = \frac{8 \pi}
{M_*^{D-2}}$ with $G_N$ and $M_*$ being the $D$-dimensional Newton 
constant and the $D$-dimensional Planck mass scale, respectively. 
Throughout this article we follow the standard 
conventions and notations of the textbook of Misner, Thorne and 
Wheeler \cite{Misner}. 

Variation of the action (\ref{1}) with respect to the $D$-dimensional
metric tensor $g_{MN}$ leads to Einstein's equations:
\begin{eqnarray}
R_{MN} - \frac{1}{2} g_{MN} R 
= - \Lambda g_{MN}  + \kappa_D^2 T_{MN},
\label{2}
\end{eqnarray}
where the energy-momentum tensor is defined as
\begin{eqnarray}
T_{MN} = - \frac{2}{\sqrt{-g}} \frac{\delta}{\delta g^{MN}}
\int d^D x \sqrt{-g} L_m.
\label{3}
\end{eqnarray}

We shall adopt the following metric ansatz:
\begin{eqnarray}
ds^2 &=& g_{MN} dx^M dx^N  \nn\\
&=& g_{\mu\nu} dx^\mu dx^\nu + \tilde{g}_{ab} dx^a dx^b  \nn\\
&=& e^{-A(r)} \hat{g}_{\mu\nu} dx^\mu dx^\nu + dr^2 
+ e^{-B(r)} d \Omega_{n-1}^2,
\label{4}
\end{eqnarray}
where $M, N, ...$ denote $D$-dimensional space-time indices, 
$\mu, \nu, ...$ do $p$-dimensional brane ones, and $a, b, ...$
do $n$-dimensional extra spatial ones, so the equality $D=p+n$
holds. (We assume $p \ge 4$.) And d$\Omega_{n-1}^2$
stands for the metric on a unit $(n-1)$-sphere, which is 
concretely expressed in terms of the angular variables $\theta_i$ as
\begin{eqnarray}
d \Omega_{n-1}^2 = d\theta_2^2 + \sin^2 \theta_2 d\theta_3^2 
+ \sin^2 \theta_2 \sin^2 \theta_3 d\theta_4^2 + \cdots
+ \prod_{i=2}^{n-1} \sin^2 \theta_i d\theta_n^2,
\label{5}
\end{eqnarray}
with the volume element $\int d \Omega_{n-1} = \frac{2 \pi^{\frac{n}
{2}}}{\Gamma(\frac{n}{2})}$.

Moreover, we shall take the ansatz for the energy-momentum tensor
respecting the spherical symmetry:
\begin{eqnarray}
T^\mu_\nu &=& \delta^\mu_\nu t_o(r),  \nn\\
T^r_r &=& t_r(r),  \nn\\
T^{\theta_2}_{\theta_2} &=& T^{\theta_3}_{\theta_3} = \cdots
= T^{\theta_n}_{\theta_n} = t_\theta(r),
\label{6}
\end{eqnarray}
where $t_i(i=o, r, \theta)$ are functions of only the radial
coordinate $r$. 

Under these ansatzs, after a straightforward calculation,
Einstein's equations reduce to
\begin{eqnarray}
e^A \hat{R} - \frac{p(n-1)}{2} A' B' - \frac{p(p-1)}{4} (A')^2
- \frac{(n-1)(n-2)}{4} (B')^2 \nn\\
+ (n-1)(n-2) e^B - 2\Lambda + 2 \kappa_D^2 t_r = 0,
\label{7}
\end{eqnarray}
\begin{eqnarray}
e^A \hat{R} + (n-2) B'' - \frac{p(n-2)}{2} A' B' 
- \frac{(n-1)(n-2)}{4} (B')^2  \nn\\
+ (n-2)(n-3) e^B + p A'' -  \frac{p(p+1)}{4} (A')^2 - 2\Lambda 
+ 2 \kappa_D^2 t_\theta = 0,
\label{8}
\end{eqnarray}
\begin{eqnarray}
\frac{p-2}{p} e^A \hat{R} + (p-1)(A'' - \frac{n-1}{2} A' B')
- \frac{p(p-1)}{4} (A')^2 \nn\\
+ (n-1) [B'' - \frac{n}{4} (B')^2 + (n-2) e^B ]  
- 2\Lambda + 2 \kappa_D^2 t_o = 0,
\label{9}
\end{eqnarray}
where the prime denotes the differentiation with respect to $r$,
and $\hat{R}$ is the scalar curvature associated with the
brane metric $\hat{g}_{\mu\nu}$.
Here we define the cosmological constant on the $(p-1)$-brane, 
$\Lambda_p$, by the equation
\begin{eqnarray}
\hat{R}_{\mu\nu} - \frac{1}{2} \hat{g}_{\mu\nu} \hat{R} 
= - \Lambda_p \hat{g}_{\mu\nu}.
\label{10}
\end{eqnarray}
In addition, the conservation law for the energy-momentum tensor,
$\nabla^M T_{MN} = 0$ takes the form
\begin{eqnarray}
t'_r = \frac{p}{2} A' (t_r - t_o) + \frac{n-1}{2} B' (t_r - t_\theta).
\label{11}
\end{eqnarray}

It is now known that there are many interesting solutions to
these equations (see, for instance, \cite{Vilenkin}).  In this article,
we shall confine ourselves to the situation where the geometry
has a warp factor, that is, 
\begin{eqnarray}
A(r) = c r,
\label{12}
\end{eqnarray}
where $c$ is a constant. Before solving a set of the equations, it
is useful to notice that for $n=1, 2$ Einstein's equations
(\ref{7}), (\ref{8}), (\ref{9}) do not include $e^B$ at all. This 
fact makes the cases of $n=1, 2$ to be quite different from
the other higher dimensional cases $n \ge 3$. Thus, in what follows,
we shall solve a set of the equations in the case of $n=2$.

\section{String-like solutions (n=2)}

Now  we shall solve the equations in the case of $n=2$. 
In this case, under the ansatz (\ref{12}), Einstein equations 
(\ref{7}), (\ref{8}), (\ref{9}) are in the form
\begin{eqnarray}
e^{cr} \hat{R} - \frac{p}{2} c B' - \frac{p(p-1)}{4} c^2
- 2\Lambda + 2 \kappa_D^2 t_r = 0,
\label{20}
\end{eqnarray}
\begin{eqnarray}
e^{cr} \hat{R}  -  \frac{p(p+1)}{4} c^2 - 2\Lambda 
+ 2 \kappa_D^2 t_\theta = 0,
\label{21}
\end{eqnarray}
\begin{eqnarray}
\frac{p-2}{p} e^{cr} \hat{R} -  \frac{p-1}{2} c B'
- \frac{p(p-1)}{4} c^2 + B'' - \frac{1}{2} (B')^2   
- 2\Lambda + 2 \kappa_D^2 t_o = 0,
\label{22}
\end{eqnarray}
and the conservation law takes the form
\begin{eqnarray}
t'_r = \frac{p}{2} c (t_r - t_o) + \frac{1}{2} B' (t_r - t_\theta).
\label{23}
\end{eqnarray}

{}From these equations, general solutions can be found as follows:
\begin{eqnarray}
ds^2 = e^{-cr} \hat{g}_{\mu\nu} dx^\mu dx^\nu + dr^2
+ e^{-B(r)} d\theta^2,
\label{24}
\end{eqnarray}
where 
\begin{eqnarray}
B(r) = cr + \frac{4}{pc} \kappa_D^2 \int^r dr (t_r - t_\theta),
\label{25}
\end{eqnarray}
\begin{eqnarray}
c^2 &=& \frac{1}{p(p+1)}(-8 \Lambda + 8 \kappa_D^2 \alpha), \nn\\
\hat{R} &=& \frac{2p}{p-2} \Lambda_p = -2 \kappa_D^2 \beta.
\label{26}
\end{eqnarray}
Here $t_\theta$ must take a definite form, which is given by
\begin{eqnarray}
t_\theta = \beta e^{cr} + \alpha,
\label{27}
\end{eqnarray}
with $\alpha$ and $\beta$ being some constants. Moreover, in
order to guarantee 
the positivity of $c^2$, $\alpha$ should satisfy an inequality
$-8 \Lambda + 8 \kappa_D^2 \alpha > 0$.

It is useful to consider two special cases of the above
general solutions. One specific solution is the one without sources
$(t_i = 0)$. Then we get a special solution found first by 
Gregory \cite{Gregory, Gherghetta}:
\begin{eqnarray}
ds^2 = e^{-cr} \hat{g}_{\mu\nu} dx^\mu dx^\nu + dr^2
+ R_0^2 e^{-cr} d\theta^2,
\label{28}
\end{eqnarray}
with $R_0$ being a constant. Here the constant $c$, the brane 
scalar curvature and the brane cosmological constant are given by
\begin{eqnarray}
c^2 &=& \frac{-8 \Lambda}{p(p+1)}, \nn\\
\hat{R} &=& \frac{2p}{p-2} \Lambda_p = 0.
\label{29}
\end{eqnarray}
In this case, as in the corresponding domain wall solution, the bulk 
geometry is the anti-de Sitter space, and the brane geometry is 
Ricci-flat with vanishing cosmological constant. 

Another specific solution occurs when 
we have the spontaneous symmetry breakdown $t_r = -t_\theta$
\cite{Vilenkin}:
\begin{eqnarray}
ds^2 = e^{-cr} \hat{g}_{\mu\nu} dx^\mu dx^\nu + dr^2
+ R_0^2 e^{-c_1 r} d\theta^2,
\label{30}
\end{eqnarray}
where
\begin{eqnarray}
c^2 &=& \frac{1}{p(p+1)}(-8 \Lambda + 8 \kappa_D^2 t_\theta) > 0, \nn\\
c_1 &=& c - \frac{8}{pc} \kappa_D^2 t_\theta, \nn\\
\hat{R} &=& \frac{2p}{p-2} \Lambda_p = 0.
\label{31}
\end{eqnarray}
Notice that this solution is more general than the previous one
(\ref{28}) since this solution reduces to (\ref{28}) when $t_\theta
= 0$.
This special solution would be utilized to analyse localization
of various matters on a string-like defect in the next section.

Finally, let us comment the solutions in general $n$. It is
straightforward to apply the above calculation procedure to
this general case, but as mentioned before the existence of the nontrivial 
terms involving $e^B$  prevent interesting solutions like
Eq.s (\ref{28}), (\ref{30}) from satisfying Einstein's equations.
We have checked that for $n \ge 3$ the solution with the warp factor
(\ref{12}) must be of the form 
\begin{eqnarray}
ds^2 = e^{-cr} \hat{g}_{\mu\nu} dx^\mu dx^\nu + dr^2 
+ R_0^2 d \Omega_{n-1}^2,
\label{32}
\end{eqnarray}
where
\begin{eqnarray}
c^2 &=& \frac{-8 \Lambda}{p(p+n-2)}, \nn\\
\hat{R} &=& \frac{2p}{p-2} \Lambda_p = 0,
\label{33}
\end{eqnarray}
where the sources satisfy the relations, $t_r + (n-1)t_\theta
-(n-2)t_o = 0$ and $t_r = t_o = constant$, which are nothing but
the relations satisfied in the spontaneous symmetry breakdown
\cite{Vilenkin}.

\section{Localization of various matters}

In this section, for clarity we shall limit our attention to a 
specific string-like solution (\ref{30}) since the generalization to 
the general solutions (\ref{24}) is straightforward.
In this paper, we have the physical setup in mind such that 
'local cosmic string' sits at the origin $r=0$ and then ask the question 
of whether variuos bulk fields with spin ranging from 0 to 2 can be
localized on the brane by means of only the gravitational interaction.
To describe 'local cosmic string' at the origin $r=0$, it is necessary
to introduce the boundary conditions meaning that the extra dimensions 
are conical around the brane with a deficit angle $\delta$, 
which are given by
\begin{eqnarray}
(e^{-\frac{1}{2}B(r)})'|^{\epsilon}_0 = - \frac{\delta}{2\pi}, 
(e^{-\frac{1}{2}B(0)})' = 1, e^{-B(\epsilon)} = 0,
\end{eqnarray}
where the boundary conditions are imposed at small radius $\epsilon$
containing the brane. But these boundary conditions are not directly
relevant to the present analysis for the localization. The only relevant
information is that the integral over the coordinate $r$ runs from 
$r=0$ to $r=\infty$, whose validity is guaranteed by the above boundary
conditions.

\subsection{Spin 0 scalar field}

In this subsection we study localization of a real scalar field
in the background geometry (\ref{30}). It will be shown 
that provided that
the constants $c$ and $t_\theta$ simultaneously satisfy certain
inequalities, there is a localized zero mode on the string-like 
defect. 

Let us consider the action of a massless real scalar coupled
to gravity:
\begin{eqnarray}
S_m = - \frac{1}{2} \int d^D x \sqrt{-g} g^{M N}
\partial_M \Phi \partial_N \Phi,
\label{34}
\end{eqnarray}
{}from which the equation of motion can be derived:
\begin{eqnarray}
\frac{1}{\sqrt{-g}} \partial_M (\sqrt{-g} g^{M N} \partial_N \Phi) = 0.
\label{35}
\end{eqnarray}
{}From now on we shall take $\hat{g}_{\mu\nu} = \eta_{\mu\nu}$
and define $P(r)=e^{-cr}$ and $Q(r)=R_0^2 e^{-c_1 r}$. 
In the background metric (\ref{30}), the equation of motion
(\ref{35}) becomes
\begin{eqnarray}
P^{-1} \eta^{\mu\nu} \partial_\mu \partial_\nu \Phi
+ P^{-\frac{p}{2}} Q^{-\frac{1}{2}} \partial_r (P^{\frac{p}{2}} 
Q^{\frac{1}{2}} \partial_r \Phi) + Q^{-1} \partial_\theta^2 \Phi
= 0.
\label{36}
\end{eqnarray}

Let us look for solutions of the form
\begin{eqnarray}
\Phi(x^M) &=& \phi(x^\mu) \chi(r) \Theta(\theta) \nn\\
&=& \phi(x^\mu) \sum_{l,m} \chi_m(r) e^{il \theta},
\label{37}
\end{eqnarray}
where the $p$-dimensional scalar field satisfies Klein-Gordon
equation
\begin{eqnarray}
\eta^{\mu\nu} \partial_\mu \partial_\nu \phi(x) = m_0^2 \phi(x).
\label{38}
\end{eqnarray}
Then Eq.(\ref{36}) reduces to
\begin{eqnarray}
\partial_r^2 \chi_m + (\frac{p}{2} \frac{P'}{P} + \frac{1}{2} \frac{Q'}{Q})
\partial_r \chi_m + (\frac{1}{P} m_0^2 - \frac{1}{Q} l^2) \chi_m = 0.
\label{39}
\end{eqnarray}
It is clear that this equation has the zero-mass ($m_0=0$) and
$s$-wave ($l=0$) constant solution $\chi_m(r) = \chi_0 = constant$.

Now we wish to show that this constant mode is localized on the
defect sitting around the origin $r=0$. The condition for
having localized $p$-dimensional scalar field is that $\chi_m(r) = \chi_0$
is normalizable. It is of importance to notice that normalizability
of the ground state wave function is equivalent to the condition that the 
"coupling" constant is nonvanishing. This key observation is fully
utilized when we discuss localization of the constant mode of 
various spin fields in the below.

Let us substitute the zero mode $\chi_m(r) = \chi_0$ into 
the starting action (\ref{34}) and check if the constant solution is 
a normalizable solution or not. With $\Phi_0(x^M) = \phi(x^\mu) \chi_0$,
the action (\ref{34}) can be cast to
\begin{eqnarray}
S_m^{(0)} &=& - \frac{1}{2} \int d^D x \sqrt{-g} g^{M N}
\partial_M \Phi_0 \partial_N \Phi_0 \nn\\
&=& - \pi \chi_0^2 \int_0^{\infty} dr P^{\frac{p}{2}-1} Q^{\frac{1}{2}}
\int d^p x \eta^{\mu\nu} \partial_\mu \phi \partial_\nu \phi. 
\label{40}
\end{eqnarray}
{}From this equation, if we define
\begin{eqnarray}
I_0 &=& \int_0^{\infty} dr P^{\frac{p}{2}-1} Q^{\frac{1}{2}} \nn\\
&=& R_0 \int_0^{\infty} dr e^{-[(\frac{p}{2}-1)c + \frac{1}{2}c_1]r},
\label{41}
\end{eqnarray}
the condition of having localized $p$-dimensional scalar field on the
defect requires that $I_0$ should be finite. Then it is easy to
rewrite this condition as inequalities for $c$ and $t_\theta$, 
whose results are given by
\begin{eqnarray}
\frac{1}{\kappa_D^2} \Lambda < t_\theta < -\frac{p-1}{2\kappa_D^2}
\Lambda,
\label{42}
\end{eqnarray}
for $c>0$ and
\begin{eqnarray}
t_\theta > -\frac{p-1}{2\kappa_D^2} \Lambda,
\label{43}
\end{eqnarray}
for $c<0$. Here for simplicity we have taken the bulk cosmological
constant $\Lambda$ to be negative, in other words, the bulk geometry
is anti-de Sitter space. (This assumption is made in this section.)
Note that the condition (\ref{42}) includes
the case without sources, (\ref{28}). This condition also implies
that the $p$-dimensional scalar field $\phi$ is localized on
the defect with the decreasing warp factor, which
exactly corresponds to localization of spin 0 field on a 
positive-tension brane in the five-dimensional 
Randall-Sundrum model \cite{Bajc}. 
We will see in subsection 4.4 that with the same condition as above,
spin 2 graviton field is localized on the defect.

\subsection{Spin 1 vector field}

It was shown in the Randall-Sundrum model in $AdS_5$ space 
that spin 1 vector field is not localized neither
on a brane with positive tension nor on a brane with negative
tension so the Dvali-Shifman mechanism \cite{Dvali} must be
invoked for the vector field localization \cite{Bajc}.
Remarkably, it will be shown in this subsection that spin
1 vector field $\it{is}$ localized on a string-like
defect, which is in sharp contrast with the domain wall case.
So we do not need to introduce additional mechanism for
the vector field localization in the case at hand.

Let us start with the action of $U(1)$ vector field:
\begin{eqnarray}
S_m = - \frac{1}{4} \int d^D x \sqrt{-g} g^{M N} g^{R S}
F_{MR} F_{NS},
\label{44}
\end{eqnarray}
where $F_{MN} = \partial_M A_N - \partial_N A_M$ as usual.
{}From this action the equation motion is given by
\begin{eqnarray}
\frac{1}{\sqrt{-g}} \partial_M (\sqrt{-g} g^{M N} g^{R S} F_{NS}) = 0.
\label{45}
\end{eqnarray}
In the background metric (\ref{30}), this equation is reduced to
\begin{eqnarray}
P^{-1} \eta^{\mu\nu} g^{M N} \partial_\mu F_{\nu N}
+ P^{-\frac{p}{2}} Q^{-\frac{1}{2}} \partial_r (P^{\frac{p}{2}} 
Q^{\frac{1}{2}} g^{M N} F_{r N}) + Q^{-1} g^{M N} 
\partial_\theta F_{\theta N} = 0.
\label{46}
\end{eqnarray}
By choosing the gauge condition $A_\theta = 0$ and decomposing
the vector field as
\begin{eqnarray}
A_{\mu}(x^M) &=& a_\mu(x^\mu) \sum_{l,m} \rho_m(r) e^{il \theta}, \nn\\
A_r(x^M) &=& a_r(x^\mu) \sum_{l,m} \rho_m(r) e^{il \theta},
\label{47}
\end{eqnarray}
it is straightforward to see that there is the $s$-wave ($l=0)$ constant
solution $\rho_m(r) = \rho_0 = constant$ and $a_r = constant$. Note that
in deriving this solution we have used $\partial_\mu a^\mu =  \partial^\mu
f_{\mu\nu} = 0$ with the definition of $f_{\mu\nu} = \partial_\mu a_\nu 
- \partial_\nu a_\mu$.

As in the scalar field in the previous subsection, 
let us substitute this constant solution into 
the action (\ref{44}) in order to see if the solution is 
a normalizable solution or not. It turns out that the action is
reduced to
\begin{eqnarray}
S_m^{(0)} &=& - \frac{1}{4} \int d^D x \sqrt{-g} g^{M N} g^{R S}
F_{MR}^{(0)} F_{NS}^{(0)} \nn\\
&=& - \frac{\pi}{2} \rho_0^2  \int_0^{\infty} dr P^{\frac{p}{2}-2} 
Q^{\frac{1}{2}} \int d^p x \eta^{\mu\nu} \eta^{\lambda\sigma} 
f_{\mu\lambda} f_{\nu\sigma}. 
\label{48}
\end{eqnarray}
Provided that we define $I_1$ by the equation
\begin{eqnarray}
I_1 &=& \int_0^{\infty} dr P^{\frac{p}{2}-2} Q^{\frac{1}{2}} \nn\\
&=& R_0 \int_0^{\infty} dr e^{-[(\frac{p}{2}-2)c + \frac{1}{2}c_1]r},
\label{49}
\end{eqnarray}
the condition of having localized $p$-dimensional vector field on the 
defect requires $I_1$ to be finite. This condition can be 
expressed as
\begin{eqnarray}
\frac{1}{\kappa_D^2} \Lambda < t_\theta < -\frac{p-3}{4\kappa_D^2}
\Lambda,
\label{50}
\end{eqnarray}
for $c>0$ and
\begin{eqnarray}
t_\theta > -\frac{p-3}{4\kappa_D^2} \Lambda,
\label{51}
\end{eqnarray}
for $c<0$. Thus, the vector field can be localized on the string-like
defect with the exponentially decreasing warp factor $c>0$
by selecting the source $t_\theta$ such that the inequality
(\ref{50}) should be satisfied. 
This result is quite different from that of the vector field
in the domain wall in $AdS_5$ \cite{Rizzo, Pomarol, Bajc} where there
was no localized solutions. The difference with the domain wall case
is that the background metric (\ref{30}) now contains the factor
$e^{-c_1 r}$ in the angular part, which gives us this nontrivial
and interesting result. Indeed, if we set $p=4$ in $I_1$, 
the contribution from $c$ vanishes and $I_1$ becomes dependent on
only $c_1$. However, this term is trivially zero in the domain wall 
case, thereby giving rise to a divergent quantity irrespective of $c$ 
in the case of the domain wall solution.
Accordingly, we do not need to invoke some additional mechanism for
localization of the vector field in the case at hand.

\subsection{Spin 1/2 fermionic field}

Now we are ready to consider spin 1/2 fermion.
Our starting
action is the Dirac action given by
\begin{eqnarray}
S_m = \int d^D x \sqrt{-g} \bar{\Psi} i \Gamma^M D_M \Psi,
\label{52}
\end{eqnarray}
{}from which the equation of motion is given by
\begin{eqnarray}
0 = \Gamma^M D_M \Psi = (\Gamma^\mu D_\mu + \Gamma^r D_r +
\Gamma^\theta D_\theta) \Psi.
\label{53}
\end{eqnarray}

We shall introduce the vielbein $e_M ^{\bar{M}}$ (and its inverse 
$e^M _{\bar{M}}$)
through the usual definition $g_{M N} = e_M ^{\bar{M}} e_N ^{\bar{N}}
\eta_{\bar{M} \bar{N}}$ where $\bar{M}, \bar{N}, \cdots$ denote
the local Lorentz indices. From the formula $\Gamma^M = e^M _{\bar{M}}
\gamma^{\bar{M}}$ with $\Gamma^M$ and $\gamma^{\bar{M}}$ being the 
curved gamma matrices and the flat gamma ones, respectively,
we have the relations:
\begin{eqnarray}
\Gamma^\mu = P^{-\frac{1}{2}} \gamma^\mu,  \ \Gamma^r = \gamma^r,
\ \Gamma^\theta = Q^{-\frac{1}{2}} \gamma^\theta.
\label{54}
\end{eqnarray}
The spin connection $\omega_M^{\bar{M} \bar{N}}$ in the covariant 
derivative $D_M \Psi = (\partial_M + \frac{1}{4} \omega_M^{\bar{M} 
\bar{N}} \gamma_{\bar{M} \bar{N}}) \Psi$ is defined as
\begin{eqnarray}
\omega_M ^{\bar{M} \bar{N}} = \frac{1}{2} e^{N \bar{M}} 
(\partial_M e_N ^{\bar{N}} - \partial_N e_M ^{\bar{N}})
- \frac{1}{2} e^{N \bar{N}} 
(\partial_M e_N ^{\bar{M}} - \partial_N e_M ^{\bar{M}}) \nn\\
- \frac{1}{2} e^{P \bar{M}} e^{Q \bar{N}}
(\partial_P e_{Q {\bar{R}}} - \partial_Q e_{P {\bar{R}}})
e^{\bar{R}} _M,
\label{55}
\end{eqnarray}
so the nonvanishing components are evaluated for the background metric 
(\ref{30}):
\begin{eqnarray}
\omega_\theta ^{\bar{r} \bar{\theta}} = - \frac{1}{2} 
Q^{-\frac{1}{2}} Q', \  \omega_\mu ^{\bar{r} \bar{\mu}} = - \frac{1}{2} 
P^{-\frac{1}{2}} P' \delta_\mu ^{\bar{\mu}}.
\label{56}
\end{eqnarray}
Therefore, the covariant derivative can be calculated to
\begin{eqnarray}
D_\mu \Psi = (\partial_\mu -  \frac{1}{4} \frac{P'}{P} \Gamma_r
\Gamma_\mu) \Psi, \  D_r \Psi = \partial_r \Psi, \  
D_\theta \Psi = (\partial_\theta -  \frac{1}{4} \frac{Q'}{Q} \Gamma_r
\Gamma_\theta) \Psi.
\label{57}
\end{eqnarray}

Substituting Eq.(\ref{57}) in the equation of motion (\ref{53}), 
we will search for the solutions of the form $\Psi(x^M) = \psi(x^\mu) 
\alpha(r) \sum e^{il \theta}$ where $\psi(x^\mu)$ satisfies 
the massless $p$-dimensional Dirac equation $\gamma^\mu 
\partial_\mu \psi = 0$. Then the equation of motion (\ref{53})
is reduced to
\begin{eqnarray}
(\partial_r +  \frac{p}{4} \frac{P'}{P} + \frac{1}{4} \frac{Q'}{Q})
\alpha(r) = 0.
\label{58}
\end{eqnarray}
The solution to this equation reads:
\begin{eqnarray}
\alpha(r) = c_2 P^{- \frac{p}{4}} Q^{- \frac{1}{4}},
\label{59}
\end{eqnarray}
with $c_2$ being an integration constant. Here we have considered
the $s$-wave solution.

Now let us show that 
the solution (\ref{59}) is normalizable if we use the
exponentially not decreasing but increasing warp factor.
\begin{eqnarray}
S_m^{(0)} &=& \int d^D x \sqrt{-g} \bar{\Psi}_0 i \Gamma^M D_M 
\Psi_0 \nn\\
&=& 2 \pi \int_0^{\infty} dr P^{\frac{p}{2} - \frac{1}{2}}
Q^{\frac{1}{2}} \alpha(r)^2 \int d^px \bar{\psi} i \gamma^\mu
\partial_\mu  \psi + \cdots.
\label{60}
\end{eqnarray}
In order to localize spin 1/2 fermion in this framework, the
integral $I_{\frac{1}{2}}$, which is defined as
\begin{eqnarray}
I_{\frac{1}{2}} &=& \int_0^{\infty} dr P^{\frac{p}{2} - \frac{1}{2}}
Q^{\frac{1}{2}} \alpha(r)^2 \nn\\
&=& c_2^2 \int_0^{\infty} dr e^{\frac{1}{2} c r},
\label{61}
\end{eqnarray}
should be finite. But this quantity is obviously divergent for 
$c>0$ while it is finite for $c<0$. This situation is the same
as in the domain wall in the Randall-Sundrum framework \cite{Bajc}
where for localization of spin 1/2 field additional localization
method by Jackiw and Rebbi \cite{Jackiw} was introduced.
This method could be applied even to the present situation, but
we believe that 
the most natural and interesting approach would be to construct a
supergravity theory corresponding to the situation at hand.

\subsection{Spin 3/2 fermionic field}

Next we turn to spin 3/2 field, in other words, the gravitino.
We will encounter the same result as in spin 1/2 field.

Let us begin with the action of the Rarita-Schwinger gravitino
field:
\begin{eqnarray}
S_m = \int d^D x \sqrt{-g} \bar{\Psi}_M i \Gamma^{[M} \Gamma^N
\Gamma^{R]} D_N \Psi_R,
\label{62}
\end{eqnarray}
{}from which the equation of motion is given by
\begin{eqnarray}
\Gamma^{[M} \Gamma^N \Gamma^{R]} D_N \Psi_R = 0.
\label{63}
\end{eqnarray}
Here the square bracket denotes the anti-symmetrization and the
covariant derivative is defined with the affine connection 
$\Gamma^R_{MN} = e^R_{\bar{M}}(\partial_M e_N^{\bar{M}} +
\omega_M^{\bar{M} \bar{N}} e_{N {\bar{N}}})$ by
\begin{eqnarray}
D_M \Psi_N = \partial_M \Psi_N - \Gamma^R_{MN} \Psi_R 
+ \frac{1}{4} \omega_M^{\bar{M} \bar{N}} \gamma_{\bar{M} \bar{N}} 
\Psi_N.
\label{64}
\end{eqnarray}
After taking the gauge condition $\Psi_\theta = 0$, the
nontrivial components of the covariant derivative are easily
calculated:
\begin{eqnarray}
D_\mu \Psi_\nu &=& \partial_\mu \Psi_\nu + \frac{1}{2} P' 
\eta_{\mu\nu} \Psi_r - \frac{1}{4} \frac{P'}{P} \Gamma_r
\Gamma_\mu \Psi_\nu, \nn\\
D_\mu \Psi_r &=& \partial_\mu \Psi_r - \frac{1}{2} \frac{P'}{P} 
\Psi_\mu - \frac{1}{4} \frac{P'}{P} \Gamma_r
\Gamma_\mu \Psi_r, \nn\\
D_r \Psi_\mu &=& \partial_r \Psi_\mu - \frac{1}{2} \frac{P'}{P} 
\Psi_\mu, \nn\\
D_r \Psi_r &=& \partial_r \Psi_r, \nn\\
D_\theta \Psi_\mu &=& \partial_\theta \Psi_\mu - \frac{1}{4} 
\frac{Q'}{Q} \Gamma_r \Gamma_\theta \Psi_\mu, \nn\\
D_\theta \Psi_r &=& \partial_\theta \Psi_r - \frac{1}{4} 
\frac{Q'}{Q} \Gamma_r \Gamma_\theta \Psi_r, \nn\\
D_\theta \Psi_\theta &=& \frac{1}{2} Q' \Psi_r.
\label{65}
\end{eqnarray}

Substituting Eq.(\ref{65}) in the equation of motion (\ref{63}), 
we will look for the solutions of the form $\Psi_\mu(x^M) = 
\psi_\mu(x^\mu) u(r) \sum e^{il \theta}, \Psi_r(x^M) = 
\psi_r(x^\mu) u(r) \sum e^{il \theta}$ where $\psi_\mu(x^\mu)$ 
satisfies the following $p$-dimensional equations $\gamma^\mu 
\psi_\mu = \partial^\mu \psi_\mu = \gamma^{[\mu} \gamma^\nu
\gamma^{\rho]} \partial_\nu \psi_\rho = 0$. 
Then the equation of motion (\ref{63}) is of the form
\begin{eqnarray}
(\partial_r +  \frac{p-2}{4} \frac{P'}{P} + \frac{1}{4} \frac{Q'}{Q})
u(r) = 0.
\label{66}
\end{eqnarray}
The solution to this equation reads:
\begin{eqnarray}
u(r) = c_3 P^{- \frac{p-2}{4}} Q^{- \frac{1}{4}},
\label{67}
\end{eqnarray}
with $c_3$ being an integration constant. Again in the above
we have considered the $s$-wave solution and $\psi_r = 0$.

We shall show that as in the case of spin 1/2 field
the solution (\ref{67}) is normalizable if we use the
exponentially increasing warp factor.
\begin{eqnarray}
S_m^{(0)} &=& \int d^D x \sqrt{-g} \bar{\Psi}_M^{(0)} i \Gamma^{[M} 
\Gamma^N \Gamma^{R]} D_N \Psi_R^{(0)} \nn\\
&=& 2 \pi \int_0^{\infty} dr P^{\frac{p}{2} - \frac{3}{2}}
Q^{\frac{1}{2}} u(r)^2 \int d^px \bar{\psi}_\mu i \gamma^{[\mu} 
\gamma^\nu \gamma^{\rho]} \partial_\nu \psi_\rho.
\label{68}
\end{eqnarray}
In order to localize spin 3/2 fermion, the
integral $I_{\frac{3}{2}}$, which is defined as
\begin{eqnarray}
I_{\frac{3}{2}} &=& \int_0^{\infty} dr P^{\frac{p}{2} - \frac{3}{2}}
Q^{\frac{1}{2}} u(r)^2 \nn\\
&=& c_3^2 \int_0^{\infty} dr e^{\frac{1}{2} c r},
\label{69}
\end{eqnarray}
must be finite. But this expression is equivalent to $I_{\frac{1}{2}}$
up to an overall constant factor so it is divergent for 
$c>0$ while it is finite for $c<0$. This result is also the same
as in the domain wall in the Randall-Sundrum framework \cite{Bajc}.

\subsection{Spin 2 field}

For the sake of completeness we briefly touch on spin 2 graviton field
from our approach since this case  has been already examined in
Ref.\cite{Gherghetta}.

Let us consider the following metric fluctuations:
\begin{eqnarray}
ds^2 &=& g_{MN} dx^M dx^N  \nn\\
&=& g_{\mu\nu} dx^\mu dx^\nu + \tilde{g}_{ab} dx^a dx^b  \nn\\
&=& e^{-cr} (\eta_{\mu\nu} + h_{\mu\nu}) dx^\mu dx^\nu + dr^2 
+ R_0^2 e^{-c_1 r} d \theta^2.
\label{70}
\end{eqnarray}
Then the equation of motion for the fluctuations $h_{\mu\nu}$
is found to be:
\begin{eqnarray}
\frac{1}{\sqrt{-g}} \partial_M (\sqrt{-g} g^{M N} \partial_N 
h_{\mu\nu}) = 0.
\label{71}
\end{eqnarray}
Consequently, it turns out that 
the equation of motion for the fluctuations in the present
background becomes equivalent to that of the scalar field considered
in subsection 4.1 \cite{Sfetsos, Csaki}. Accordingly, 
we expect that the condition
for localization of spin 2 field might be equivalent to that of spin
0 field. This is indeed the case as shown in what follows.

Let us look for solutions of the form
\begin{eqnarray}
h_{\mu\nu}(x^M) &=& \check{h}_{\mu\nu}(x^\mu) \varphi(r) 
\Theta(\theta) \nn\\
&=& \check{h}_{\mu\nu}(x^\mu) \sum_{l,m} \varphi_m(r) e^{il \theta},
\label{72}
\end{eqnarray}
where 
$\eta^{\mu\nu} \partial_\mu \partial_\nu \check{h}_{\rho\sigma}
 = m_0^2 \check{h}_{\rho\sigma}$.
It is then easy to show that the equation of motion has the zero-mass 
($m_0=0$) and $s$-wave ($l=0$) constant solution $\varphi_m(r) = 
\varphi_0 = constant$.
Substitution of this zero mode $\varphi_m(r) = \varphi_0$ into 
the Einstein-Hilbert action (\ref{1}) leads to 
\begin{eqnarray}
S^{(0)} \sim  \varphi_0^2 \int_0^{\infty} dr P^{\frac{p}{2}-1} 
Q^{\frac{1}{2}} \int d^p x \partial^\rho \check{h}^{\mu\nu}
\partial_\rho \check{h}_{\mu\nu} + \cdots. 
\label{73}
\end{eqnarray}
{}From this equation, if we define $I_2$ by
\begin{eqnarray}
I_2 = \int_0^{\infty} dr P^{\frac{p}{2}-1} Q^{\frac{1}{2}},
\label{74}
\end{eqnarray}
the condition of having localized $p$-dimensional graviton field on the 
defect requires that $I_2$ should be finite. Note that $I_2 = I_0$,
so that we have the same result for localization of the graviton
as in the spin 0 scalar field.

\section{Discussions}

In this paper, we have investigated  two problems, those are,
finding solutions with the warp factor corresponding to string-like 
defects and checking localization of various spin 
fields on such a string-like defect from the viewpoint of field
theory.
We have presented more general solutions compared to the solutions
found so far. Moreover, it has been 
found that spin 0 and 2 fields are localized on a defect with 
the exponentially decreasing warp factor, and spin 1 field can be also 
localized on a defect with the exponentially decreasing warp factor 
by selecting an appropriate range of values of sources.
On the contrary, spin 1/2 and 3/2 fields can be 
localized on a defect with the exponentially increasing warp 
factor. 

These results for localization of various spin fields
coincide with the corresponding ones \cite{Bajc}
in the Randall-Sundrum model \cite{Randall2} and many brane 
model \cite{Oda1, Oda2} except spin 1 vector field.
It is remarkable that there is no localized vector field
on the brane in the domain wall model, whereas vector
field can be localized on the defect in the string-like
model owing to the existence of the nontrivial exponential
factor in the angular part of the metric.

Localizing the fermionic degrees of freedom on the brane
or the defect requires us to introduce other interactions but
gravity. The most natural approach for it seems to embed the present
model in the supergravity theory. In this respect, it is worthwhile
to emphasize that six-dimensional supergravity is more beautiful than
five-dimensional supergravity. Recently, the authors in Ref.\cite{Mario}
have studied a covariant formalism of six-dimensional supergravity.
In future, we wish to extend the present model to the supergravity
model.

\vs 1

\end{document}